\documentclass[twoside,english,letterpaper,english,aps,pra,twocolumn,showpacs,floatfix]{revtex4}

\usepackage{babel}
\usepackage[latin9]{inputenc}
\usepackage{color}
\usepackage{verbatim}
\usepackage{amsmath}
\usepackage{graphicx}

\makeatletter

\providecommand{\tabularnewline}{\\}

\@ifundefined{textcolor}
{%
 \definecolor{BLACK}{gray}{0}
 \definecolor{WHITE}{gray}{1}
 \definecolor{RED}{rgb}{1,0,0}
 \definecolor{GREEN}{rgb}{0,1,0}
 \definecolor{BLUE}{rgb}{0,0,1}
 \definecolor{CYAN}{cmyk}{1,0,0,0}
 \definecolor{MAGENTA}{cmyk}{0,1,0,0}
 \definecolor{YELLOW}{cmyk}{0,0,1,0}
 }

\makeatother

\begin{document}

\title{Reducing sequencing complexity in dynamical quantum error suppression\\
by Walsh modulation}

\author{David Hayes}
\affiliation{Joint Quantum Institute, University of Maryland, College
Park, MD 20742 USA}

\author{Kaveh Khodjasteh}
\affiliation{Department of Physics and Astronomy, Dartmouth College,
Hanover, NH 03755 USA}

\author{Lorenza Viola}
\affiliation{Department of Physics and Astronomy, Dartmouth College,
Hanover, NH 03755 USA}

\author{Michael J. Biercuk}
\email{michael.biercuk@sydney.edu.au}
\affiliation{School of Physics and Centre for Engineered Quantum Systems, 
The University of Sydney, NSW 2006 Australia}

\date{\today}


\begin{abstract}
We study dynamical error suppression from the perspective of reducing sequencing complexity, in order to facilitate efficient semi-autonomous quantum-coherent systems.  With this aim, we focus on  \emph{digital} sequences where all interpulse time periods are integer multiples of a
minimum clock period and compatibility with simple digital classical control circuitry is intrinsic, using so-called {\em Walsh functions} as a general mathematical framework.  The Walsh functions are an orthonormal set of basis functions which may be associated directly with the control propagator for a digital modulation scheme, and
dynamical decoupling (DD) sequences can be derived from the locations
of digital transitions therein.  We characterize the suite of the resulting {\em Walsh dynamical decoupling} (WDD) 
sequences, and identify the number of periodic square-wave (Rademacher) functions required to generate a Walsh function as the key determinant of the error-suppressing features of the relevant WDD sequence.  WDD forms a unifying theoretical framework as it includes a large variety of well-known and novel DD sequences, providing significant flexibility and performance benefits relative to basic quasi-periodic design.
We also show how Walsh modulation may be employed for the
protection of certain nontrivial logic gates, providing
an implementation of a dynamically corrected gate.  Based on 
these insights
we identify Walsh modulation as a digital-efficient approach for physical-layer
error suppression.
\end{abstract}

\maketitle

\section{Introduction}

Dynamical quantum error correction has been proposed as a strategy by
which arbitrarily
accurate evolutions may in principle be implemented in a large class
of open quantum systems.  This approach involves application of open-loop control
protocols at the physical
level~\cite{Viola1998,Viola1999,Zanardi1999,Vitali1999,Viola2003,Byrd2003,Kofman2004,Khodjasteh2005,Yao2007,Uhrig2007,Gordon2008,khodjasteh2009dcg,khodjasteh2009pradcg,khodjasteh2010cdcg,Liu2010,Biercuk_Filter};
through time-dependent modulation of the system's
dynamics, the effects of an environment which fluctuates sufficiently slowly are coherently averaged out. Dynamical decoupling (DD) is
an experimentally
validated~\cite{Ladd2005,Fraval2005,Krojanski2006,Biercuk2009,BiercukPRA2009,Du2009,West2009,Damodarakurup2009,Szwer2011,Alvarez2010,deLange2010,Sagi2010,Bluhm_LongCoherence,Barthel2010,Cory2010,Oliver_PC,Medford2011}
subclass of these protocols specifically tailored to the task of
suppressing decoherence during the implementation of the identity
operator -- resulting in improved quantum storage.

DD takes physical inspiration from the spin echo in nuclear magnetic
resonance (NMR)~\cite{Hahn50,Haeberlen1976,Vandersypen2004}, and
relies on deterministic (periodic~\cite{Viola1998} or
aperiodic~\cite{Khodjasteh2005,Uhrig2007}), or even
random~\cite{Viola2005,Viola2006} modulation of the idle system via
pulsed control. A proliferation of analytical formalisms and new DD
schemes has appeared in the literature, in particular for the
paradigmatic case of a single qubit exposed to pure (classical and/or
quantum)
dephasing~\cite{Viola1998,Faoro2004,Uhrig2007,Uhrig2008,Kuopanporttii2008,West2009,Biercuk2009,Uys2009,Hodgson2010,Viola:BADD}.
While an understanding of the performance of these sequences may be
unified by the application of a \emph{noise filtering}
framework~\cite{Martinis2003,Kofman2004,Cywinski2008,Suter_Filter,Biercuk_Filter},
each sequence brings particular requirements for the necessary pulse
timings, often incorporating nonintuitive analytical expressions or
numerical search to define pulse locations in a sequence. As a result,
the generation of DD pulse sequences at the lab bench is usually
accomplished using specially programmed microcontrollers or a PC under
user control.

In this work, we address the problem of control complexity in
dynamical error correction, introducing a set of \emph{digital} DD protocols
optimized for hardware compatibility and minimization of sequencing
complexity.  These protocols are based on the \emph{Walsh
functions}~\cite{Walsh_Beauchamp,Walsh_Tzafestas}, which take binary
values and are composed of products of square waves, forming an
orthonormal basis similar to the sines and cosines.  The Walsh
functions benefit from compact notation and a uniform mathematical
basis for sequence construction.
We describe the error-suppressing properties of Walsh modulation and Walsh dynamical decoupling (WDD), and introduce a quantitative metric for
\emph{sequencing complexity}, $r$, the number of Rademacher
square-wave functions which must be multiplied (or added mod-2) in
hardware to generate the control propagator for that sequence.  The
Rademacher functions in turn can be trivially generated via elementary
digital circuits synchronous with a distributed clock signal.  Thus, by construction Walsh modulation is highly compatible with simple digital sequencing circuitry and digital clocking as may be needed in large or semi-autonomous systems.  This approach therefore provides {\em efficient} error suppression,
important for real-world implementations beyond simple demonstration experiments.

The remainder of this manuscript is organized as follows. After
introducing the relevant system and control setting in
Section~\ref{Sec:Setting}, Section~\ref{Sec:Walsh} is devoted to
describing the mathematical formulation and key features of the Walsh
functions, and their natural use in defining WDD schemes.  In this
section, our main results are established, including an exact
relationship between the sequencing complexity $r$ and the order of
decoherence suppression in the perturbative limit.
Here, we also describe the entire taxonomy of WDD sequences,
discussing relationships between WDD and well-known sequences, as well
as characterizing new DD sequences with digital timing.  In
Sec.~\ref{Sec:Beyond}, we discuss extensions of the WDD formalism
beyond the simplest setting of a single qubit exposed to classical
dephasing noise.  In particular, we show how two-axis generalizations
of WDD naturally recover and expand existing concatenated DD schemes,
and discuss how Walsh modulation allows enhancement of the fidelity of nontrivial gate operations, examining a recent trapped-ion experiment as an example
\cite{Hayes2011}.  We follow this with an analysis of the explicit
benefits of WDD over other optimized DD approaches in
Sec.~\ref{Sec:Benefits}, and close with a brief summary and outlook.
A proof of WDD error suppression properties from Rademacher functions
is included in Appendix A for completeness.

\section{Physical Setting}
\label{Sec:Setting} 

DD is applicable to a variety of non-Markovian error models (including
arbitrary many-qubit systems interacting with quantum environments,
\cite{Viola1999,Liu2011}) and can address non-ideal pulses or
continuous control scenarios (as in Eulerian DD \cite{Viola2003}).
Our starting point here, however, is the simplest yet practically
important case of a single qubit subject to classical phase noise and
controlled with ideal $\pi$ pulses. In this setting, the qubit is
affected by an undesired noise Hamiltonian,
\begin{equation}
H_{\text{noise}}=\beta(t)\sigma_{z},
\label{eq:Hn}
\end{equation}
where $\beta(t)$ is a stochastic process and $\sigma_{i}$ denote the
Pauli matrices for $i=x,y,z$. We assume that the system can be
controlled by application of idealized sequences of qubit rotations,
each corresponding to a $\pi$ rotation around the $x$ axis, i.e., to
the unitary operator $\sigma_{x}$. Each sequence is characterized by
the pulse timings $\{t_{j}\}_{j=1}^{s}$, 
where $t_{0}=0$ and $t_{s+1}=\tau$ denote the initial (preparation)
and final (readout) times respectively.  We use
$\delta_{j}=t_{j}/\tau$ to denote the ``normalized pulse locations''
and the ``pulse pattern'' $p=\{\delta_{j}\}_{j=1}^{s}$, to
distinguish among various sequences with the same running time
\cite{note0}.

During the evolution, the $\pi$ pulses implement the control
propagator
\begin{eqnarray*}
U_{c}(t) & = & \sigma_{x}^{[y(t)+1]/2} , 
\end{eqnarray*}
where $y(t)$ takes values $\pm 1$ and switches instantaneously between
these values at times corresponding to application of the $\pi$
pulses. For brevity, we shall refer to $y(t)$ as the ``sequence
propagator'' in what follows. The ``filter function''
$F_{p}(\omega\tau)$ associated with the pulse pattern
$p=\{\delta_{j}\}$ and sequence duration $\tau$ is defined in terms of $\tilde{y}(\omega\tau)$,
the Fourier transform of the sequence
propagator \cite{Cywinski2008,Uhrig2008}:
\begin{align}
F_{p}(\omega\tau) & =\omega^{2}|\tilde{y}(\omega\tau)|^{2}
\label{eq:fff}\\
& =\Big\vert\sum\limits _{j=0}^{s}(-1)^{j}
(e^{i\delta_{j}\omega\tau}-e^{i\delta_{j+1}\omega\tau})\Big\vert^{2}.
\label{eq:ff}
\end{align}
The case of free evolution (also referred to as free induction decay,
FID, in NMR terminology) is formally included by letting $s=0$.

The presence of the noise Hamiltonian (\ref{eq:Hn}) implies a
coherence loss at the readout due to phase randomization, resulting
from the ensemble average with respect to noise realizations.  The
action of DD is to break the system's evolution into a sequence of
interactions with the environment with \emph{alternating signs},
resulting in significantly reduced phase accumulation at the end of
the sequence \cite{Viola1998,Viola1999}.  This may be effectively
represented as the expected value of the convolution of the stochastic
noise term with the sequence propagator
\cite{Haeberlen1976,Uhrig2007,Uhrig2008,Cywinski2008,Biercuk_Filter}.
In particular, the filter function provides a compact {\em exact}
expression for the coherence decay under Gaussian noise; if the system
is prepared in a superposition of eigenstates of $\sigma_{z}$, its
coherence $W\equiv |\overline{\langle\sigma_{+}\rangle(\tau)}|$ decays
as $e^{-\chi_{p}(\tau)}$, where
\begin{equation}
\chi_{p}(\tau)=\frac{2}{\pi}\int\limits _{0}^{\infty}
\frac{S_{\beta}(\omega)}{\omega^{2}}F_{p}(\omega\tau)d\omega,
\label{Eq:chi}
\end{equation}
and $S_{\beta}(\omega)$ is the power spectrum of the noise
$\beta(t)$. While nominally the integration range in
Eq. (\ref{Eq:chi}) is infinite, in practice $S_{\beta}(\omega)$, and
thus the ``spectral measure'' $\lambda (\omega) \equiv
S_{\beta}(\omega)/2\pi \omega^2$ \cite{Viola:BADD}, is significant
only for frequencies smaller than an ultraviolet cut-off frequency
$\omega_{c}$.  The filter function can also be employed for a
qubit coupled to a purely-dephasing quantum bosonic environment with a
similar expression for decoherence under DD sequences
\cite{Viola1998,Uhrig2007}.  The effect of continuous control can also
be approximated within the filter function formalism as long as terms
that are of second order and higher in $H_{\text{noise}}$ are
ignored \cite{Kofman2004,Gordon2008} (see also Sec. \ref{Sec:Beyond}).

The filter function enters the integrand in
Eq. (\ref{Eq:chi}) as a multiplicative factor of $\lambda(\omega)$,
therefore, {\em as long as $F_{p}(\omega\tau)$ is small for}
$\omega<\omega_{c}$, the coherence loss will also remain small, as
desired. Determining the actual value of $\chi_p(\tau)$ requires a detailed knowledge of
$S_{\beta}(\omega)$, yet we may compare the coherence associated with
various DD sequences (including free evolution) by directly comparing
their corresponding filter functions over the interval
$[0,\omega_{c}\tau${]}.  For all sequences the filter function vanishes at zero, however we may
differentiate the low-frequency behaviors (rates of growth) among various
timing patterns.  Let the low-frequency behavior of the filter
function be given by
\begin{equation}
F_{p}(\omega\tau)\propto(\omega\tau)^{2(\alpha+1)},
\label{eq:powlaw}
\end{equation}
corresponding to 
\[
\tilde{y}(\omega\tau)\propto(\omega\tau)^{\alpha}.
\]
In the language of filter design \cite{Biercuk_Filter},
Eq. (\ref{eq:powlaw}), defines a ``highpass filter'' with a
``rolloff'' of $6(\alpha+1)\text{dB}/\text{octave}$.  As long as the
cutoff frequency $\omega_{c}$ is sufficiently small, the low-frequency
behavior of the filter function translates to
$$1-W\propto(\omega_{c}\tau)^{2\alpha+1}.$$  
\noindent 
For example, the filter function for free evolution
$F_{\{0,1\}}(\omega\tau)=2\sin^{2}(\omega\tau/2)$ corresponds to
$\alpha=0$, while high-order DD sequences have larger positive
$\alpha$'s. The frequency range over which the filter function
suppresses noise is given by the ``bandwidth'' $\Omega_{p}$, roughly
defined as the largest frequency 
below which $F_{p}(\omega\tau)\le1$, a value approximately
commensurate with the value $\omega_{F1}$ introduced in previous
analyses of the filter function~\cite{Biercuk_Filter}.  Large values
of the bandwidth $\Omega_{p}$ improve the high-frequency robustness of
a particular DD sequence and allow it to be used with comparatively
lower pulse rates.

\section{Dynamical Decoupling by Digital Modulation}
\label{Sec:Walsh}

Many previously developed schemes for DD rely on access to sequences
defined in continuous time, which is a good approximation for most
benchtop experiments conducted today.  In the long term, however, where large systems with error rates deep below the fault-tolerance threshold are required,
these approximations will cease to provide an accurate estimate of
residual error rates.   Previous work has shown that the benefits of
optimized sequences such as UDD become diminished in such cases where we expect the use of digital control and discretized time~\cite{Biercuk_Filter}.  We are thus
motivated to find \emph{digital modulation} schemes that are both
intrinsically compatible with discrete time and with the digital
control hardware that will inevitably be employed in sequencing.

We define two relevant timescales for such a modulation scheme; the
total running time $\tau$, and a \emph{minimum interval} (minimum
switching time \cite{Viola:BADD}) given by
$$\tau_{\text{min}}=\tau/2^m,$$
\noindent
where $2^m$ is the largest possible number of 
free-evolution periods in an applied sequence.  In practice,
$\tau_{\text{min}}$ is bounded from below by technological constraints
such as modulation rates or hardware clock speeds.  In the case of
digital modulation, all interpulse periods must be defined as integer
multiples of $\tau_{\text{min}}$.  This in turn places constraints on
the allowed values of ${\delta_{j}}$ in a sequence.  Irrational values
of fractional pulse locations have intrinsic conflict with digital
modulation, thus mandating alternate approaches.

In order to overcome this challenge, we have identified the Walsh
functions as a mathematical basis that is compatible with digital
sequencing hardware.  In this section we discuss the Walsh functions
and the WDD derived from them.

\subsection{Walsh functions and WDD}
\label{Sub:WDD}
\begin{figure*}[htb]
\begin{centering}
\includegraphics[width=\textwidth]{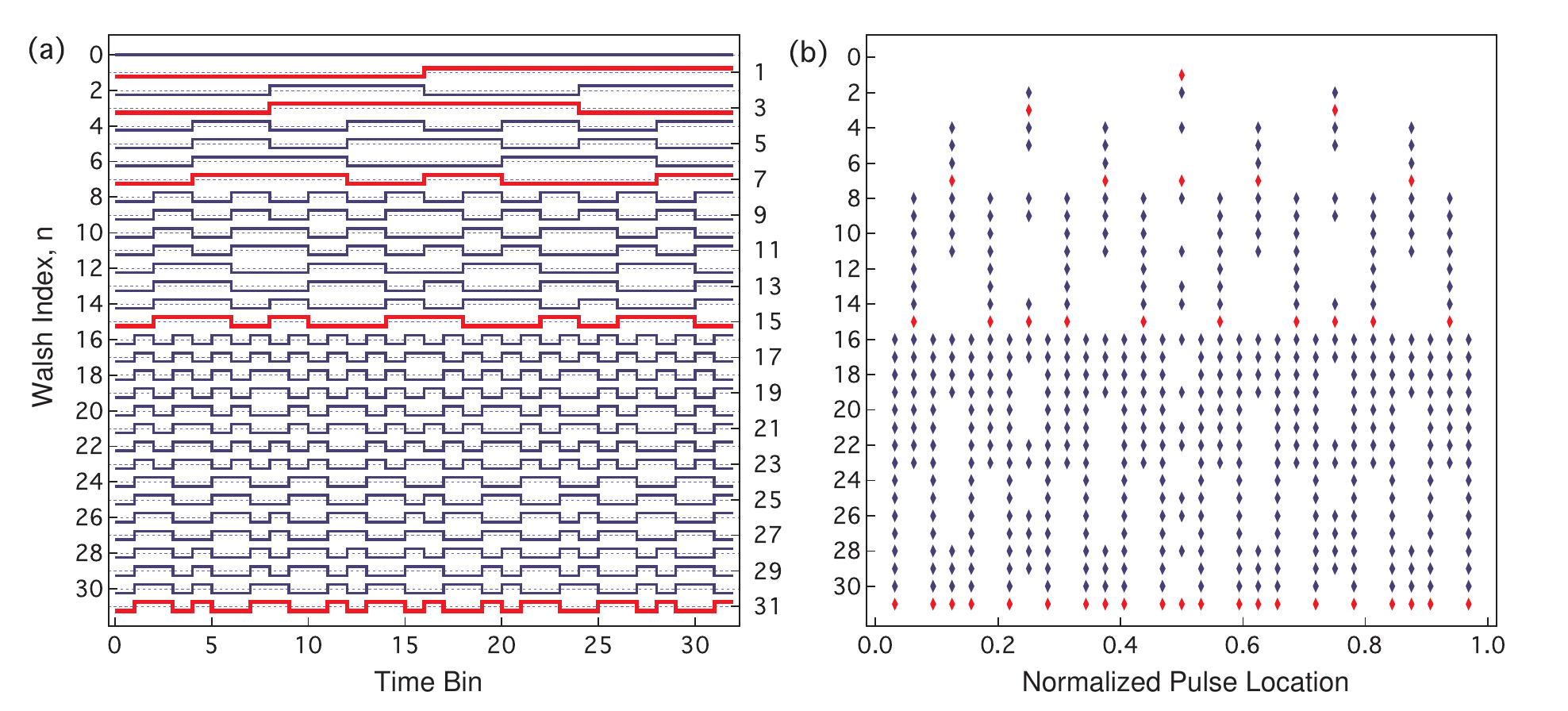}\\
\caption{(color online) Walsh functions listed by Paley ordering. (a) The first 32
Walsh functions.  (b) The normalized pulse locations, $\delta_{j}$,
corresponding to the digital transition points of Walsh functions.  Sequences
$\text{WDD}_{2^{r}-1}$ are highlighted in both panels.
\label{Fig:F1} } \par\end{centering}
\centering
\end{figure*}
The Walsh functions are a family of binary valued ($\pm1$)
piecewise-constant functions on the $[0,1]$ interval
\cite{walsh1923}. They found a place in engineering in the 1960s, when
they started being applied to problems ranging from communications and
signal analysis to image processing, and noise filtering
\cite{Walsh_Beauchamp,Walsh_Tzafestas}.  The Walsh functions come in a
variety of labeling conventions, including the Hadamard,
``sequency'', and Paley (or dyadic). In particular, the sequency
ordering counts the number of ``switchings'' of the Walsh
functions. In this paper, however, we focus on the Paley ordering,
given in terms of the so-called {\em Rademacher functions}
\cite{rademacher1922}, which are defined as
\[
\textsf{R}_{j}(x)=\textrm{sgn}\left[\sin\left(2^{j}\pi
x\right)\right],\;\; j \geq 0, 
\]
and correspond to periodic digital switchings between $\pm1$ over
$[0,1]$ with the ``rate'' $2^{j}$. The Walsh function of Paley order
$n$, $\textsf{W}_{n}(x)$, is then defined as \cite{Paley1932}
\begin{eqnarray}
\textsf{W}_{n}(x) & = & \prod_{j=1}^{m}\textsf{R}_{j}(x)^{b_{j}}
,\;\;\; x\in [0,1],
\label{eq:walnrad}
\end{eqnarray}
where we denote by $(b_{m}b_{m-1}...b_{1})_{2}$ the \emph{binary}
representation of $n$, that is, $n=b_m 2^{m-1} + b_{m-1} 2^{m-2}
+\ldots + b_1 2^0$. The actual number, $r$, of Rademacher functions
used in constructing a Walsh function, is the {\em Hamming weight}
(number of non-zero binary digits) of $n$. For illustration, the Walsh
functions $\{\textsf{W}_{n}(x)\}_{n=1}^{32}$ in the Paley ordering are
shown in Fig.~\ref{Fig:F1}(a).  The Walsh functions form an
orthonormal basis over $[0,1]$, that is,
\begin{equation}
\int_{0}^{1}\textsf{W}_{n}(x)\textsf{W}_{m}(x)dx=\delta_{mn} , 
\end{equation}
where $\delta_{mn}$ is the Kronecker delta. Any integrable function
$f(x)$ defined over $[0,1]$ has a convergent Walsh-Fourier expansion
similar to the usual Fourier series:
\begin{equation}
f(x)=\sum_{n=0}^{\infty}a_{n}\textsf{W}_{n}(x), \;\;
a_{n}=\int_{0}^{1}f(x)\textsf{W}_{n}(x)dx.
\label{Eq:Synthesis}
\end{equation}
This allows us to expand any sequence propagator in the Walsh
basis. We note in passing that the Walsh basis is technically
over-complete while the Rademacher functions form an orthonormal
system \cite{rademacher1922}.

We define the {\em Walsh DD} (WDD) sequences directly based on the
Walsh functions: that is, for a given running time $\tau$, the
sequence $\text{WDD}_{n}$ is determined by the control propagator
\begin{equation}
y(t)=\textsf{W}_{n}(t/\tau), \;\; t\in[0,\tau].
\label{eq:yWal}
\end{equation}
Equivalently, we may specify the normalized pulse locations for
$\text{WDD}_{n}$ as the switching locations of $\textsf{W}_{n}(x)$,
whereas the constant pieces of the Walsh functions correspond to
interpulse delays. By construction, the number of pulses used in
$\text{WDD}_{n}$ is given by the sequency $s$ associated with
$\textsf{W}_{n}$. More explicitly, if in binary notation
$n=(b_{m}b_{m-1}...b_{1})_{2}$ and $s=(g_{m}g_{m-1}...g_{1})_{2}$,
then $g_{i}=b_{i}+b_{i+1}\text{mod 2}$ \cite{note2}, known as a Gray
code.

Both the total running time, $\tau$, and the minimum switching time,
$\tau_{\text{min}}=\tau/2^{m}$, impose constraints on the accessible
WDD sequences.  First, we must have $m=O(\log_{2}n)$, the number of
digits in the binary representation of $n$ for a given WDD sequence.
Once these parameters are fixed, there is a maximum value $n\approx
O(\tau/\tau_{\text{min}})$ for which $\text{WDD}_{n}$ is
viable. Reversing this argument, by focusing on a finite set of
$\text{WDD}_{n}$ and a fixed running time, we automatically
accommodate a finite minimum switching time.

Figure \ref{Fig:F1}(b) depicts the pulse locations plotted as
$\delta_{j}$ for a normalized sequence duration $\tau=1$ extracted
from the $\textsf{W}_{32}(t/\tau)$.  Some familiar DD sequences are
immediately identified: $\text{WDD}_{1}$ is the spin echo sequence and
$\textrm{WDD}_{3}$ is the two-pulse Carr-Purcell-Meiboom-Gill (CPMG)
pulse sequence.  We will return to these correspondences in Section~\ref{subsec:classification}.

\subsection{Error suppression properties}
\label{subsec:Rademacher}

A primary aim in the construction of DD sequences is to increase the
order of error suppression, resulting in better cancellation of noise
at sufficiently low
frequencies~\cite{Uhrig2007,Uhrig2008,Lee2008,Cywinski2008,Yang2008}. In
order to characterize the error suppression capabilities of WDD
sequences, we begin by noting that the Walsh functions, constructed as
products of $r$ Rademacher functions {[}Eq. (\ref{eq:walnrad}){]}
satisfy the following important equality (see Appendix for an explicit
proof):
\begin{equation}
\int_{0}^{1}[\textsf{R}_{j_{1}}(x) \cdots\textsf{R}_{j_{r}}(x)]x^{k}dx
\equiv 0, \;\;\; k=0,\dots,r-1 ,
\label{eq:Rad:int}
\end{equation}
where the $\{j_{k}\}$ indices label the locations of the $r$ non-zero
digits in the binary representation of $n$.  Equivalently, the
monomials $x^{k}$ have {\em no} Walsh-Fourier component $a_{n}$ as
long as $n$ has a Hamming weigth of at least $k$.  Let us focus on the
Fourier transform of the propagator function for
$\text{WDD}_{n}$. Using Eqs. (\ref{eq:fff}) and (\ref{eq:yWal}) we
have:
\begin{align}
\tilde{y}(\omega\tau)= & \:
\tau\int_{0}^{1}\textsf{W}_{n}(x)e^{i\omega\tau x}dx , \nonumber \\ =
& \: \tau\int_{0}^{1}\textsf{R}_{j_{1}}(x)\cdots
\textsf{R}_{j_{r}}(x)\sum_{k=r}^{\infty}\frac{(i\omega\tau
x)^{k}}{k!}dx , \nonumber
\end{align}
where the powers of $x$ below $r$ have been eliminated by using
Eq.~(\ref{eq:Rad:int}).  Together with Eq. (\ref{eq:fff}), the above
equation implies that
\begin{equation}
F_{\text{WDD}_{n}}(\omega\tau)\propto(\omega\tau)^{2(r+1)}.
\label{fwdd} 
\end{equation} 
Thus, $\text{WDD}_{n}$ {\em suppresses errors up to order $r$}.
Expressed differently, all $\text{WDD}_{n}$ derived from Walsh
functions composed of $r$ Rademacher functions exhibit the same order
of error suppression.  The relationship between $r$, the Hamming
weight of the Walsh order $n$, and the order of error suppression is
one of the main results of this work.

The error-suppressing properties of WDD$_n$ can also be directly
established upon obtaining a compact expression for the corresponding
filter function (in analogy to \cite{Uhrig2008}), which is possible by
using Eq. (\ref{eq:ff}) in combination with a change of variable
$e^{i\omega\tau/2^{m}}\mapsto z$.  For $n=(b_{m}\cdots b_{0})_{1}$,
the filter function for $\text{WDD}_{n}$ is given by
\begin{align}
\hspace*{-3mm}F_{\text{WDD}_{n}}(\omega\tau) &
 =\Big\vert(1-z)\prod_{j=1}^{m}(1+(-1)^{b_{j}}z^{2^{j-1}})
 \Big\vert^{2}\nonumber \\ &
 =4^{m+1}\sin^{2}(\omega\tau_{\text{min}}/2) \nonumber \\
 \times\!\prod_{\{j\vert
 b_{j}=1\}}\!\!\!\!\sin^{2}(2^{j-2}\omega\tau_{\text{min}}) &
 \times\!\prod_{\{j\vert
 b_{j}=0\}}\!\!\!\!\cos^{2}(2^{j-2}\omega\tau_{\text{min}}).
\label{eq:formula}
\end{align}
where the product chooses $\sin$ or $\cos$ factors based on the $j$-th
binary digit $b_j$ of $n$.  Interestingly, the trigonometric factors
appearing in Eq.~(\ref{eq:formula}) have complementary implications
for the filter function and ultimately for error suppression.  The
smallest common period among these terms is at
$\omega\tau_{\text{min}}=2\pi$. Each sine term is linear in
$\omega\tau$ at $\omega=0$ and each factor of
$\sin^{2}(2^{j-2}\omega\tau_{\text{min}})$ subsequently contributes a
factor proportional to $(\omega\tau)^{2}$ to the filter function.
These combine to give
$F_{\text{WDD}_{n}}(\omega\tau)\propto(\omega\tau)^{2(r+1)}$, consistent
with Eq. (\ref{fwdd}).  Also, all sine terms have zeroes at
$\omega\tau_{\text{min}}=k\pi$ for integer $k$, that correspond to
zeroes at $\omega=2^{m}k\pi/\tau$.  In contrast, they all carry large
maxima occurring at $\omega=O(\tau_{\text{min}}^{-1})$.  The cosine
terms on the other hand have no effect on the filter function around
$\omega=0$, but drop to zero at somewhat higher frequencies (at which
the sine terms might be significantly large).  This suggests that they
can reduce the spikes of the sine terms and contribute to increasing
the bandwidth $\Omega_{\text{p}}$.

\begin{figure*}
\includegraphics[width=15cm]{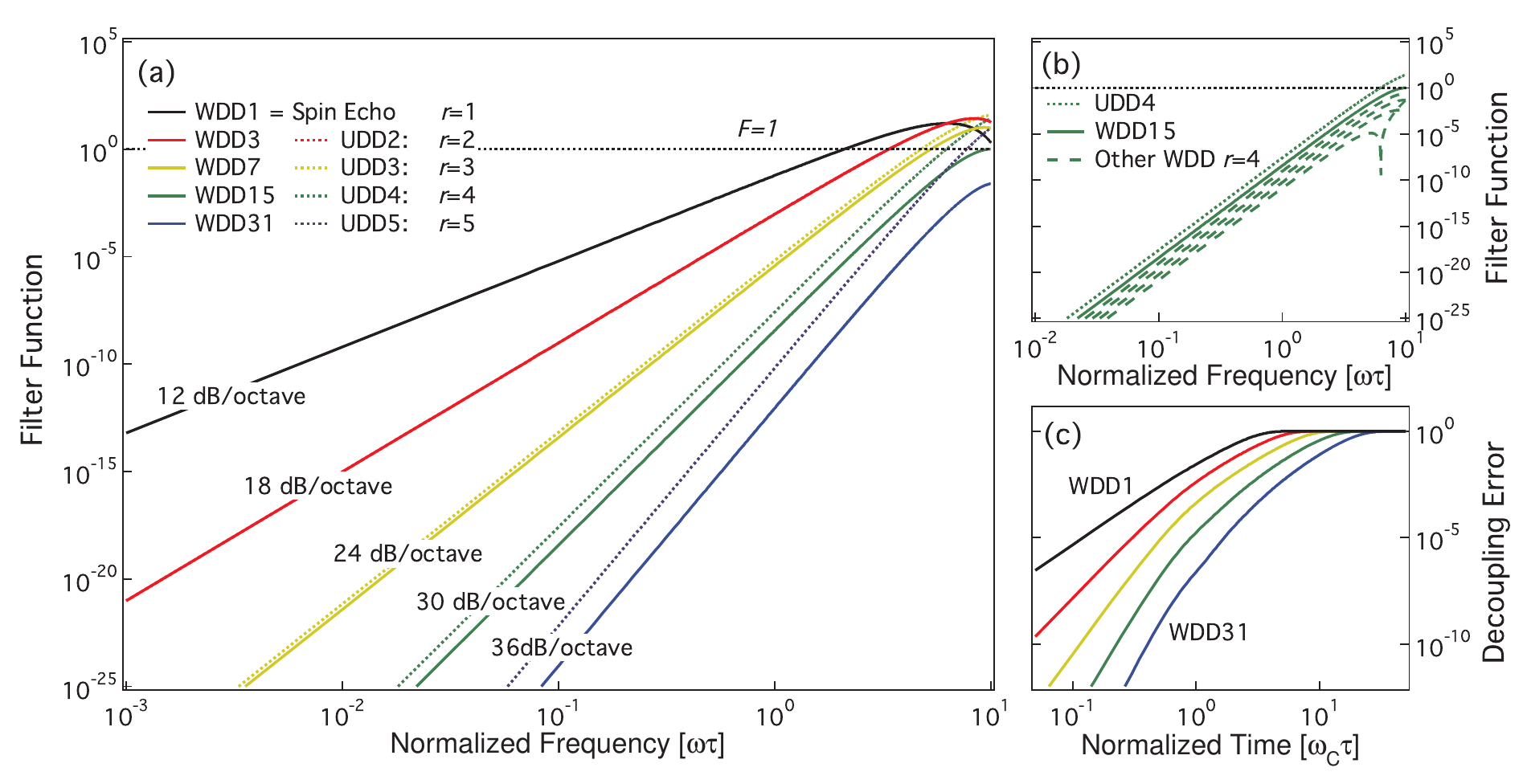}\\
\caption{(color online) Error suppression properties of WDD sequences. (a) Filter
functions calculated for sequences derived from $W_{2^{r}-1}(t/\tau)$
and corresponding to CDD$_{r}$, with fixed total sequence duration,
$\tau$. For each value of $r$, the number of fundamental Rademacher
functions that are needed to produce the sequence increases, and the
order of error-suppression, demonstrated graphically by the
low-frequency rolloff of the filter function, increases by 6
dB/octave. Dashed lines correspond to the UDD sequence providing the
same order of error suppression.  (b) Filter functions for all WDD
sequences with $n<32$ and $r=4$.  UDD4 is represented by a dashed line
and WDD15 by a solid line.  Dashed lines correspond, left-to-right, to
WDD23, WDD27, WDD29, WDD30. (c) Decoupling error as a function of
dimensionless frequency normalized by the cutoff frequency,
$\omega_{c}$ in $S_{\beta}(\omega)=\alpha
\omega^{-2}\exp{[-(\omega/\omega_{c})^{2}]}$.  Here,
$\alpha=5\times10^{16}$, and frequency ranges are set as in
Ref.~\cite{BiercukPRB2011}.  Sequences are same as in (a), and increase in $n$ from left to right. \label{Fig:F2} } 
\end{figure*}

With the insights above, we find that sequences $\text{WDD}_{2^{r}-1}$
{\em have the lowest value of sequency} (pulse number) for a given
order of error suppression (among the Walsh family).  This is because the binary representation
of a digital integer $n=2^{r}-1$ requires all $b_{i}=1$. Figure
\ref{Fig:F2} depicts the filter functions (in a log-log scale)
associated with $\text{WDD}_{2^{r}-1}$ for $r=1,\cdots,5$ (as
identified in Fig.~\ref{Fig:F1}), where we illustrate how the
low-frequency rolloff is increased by increasing $r$.  We compare the
filter functions for these sequences to those of Uhrig DD (UDD), known
to provide $s$-order error suppression given $s$ pulses
\cite{Uhrig2007,Uhrig2008}.

In Fig.~\ref{Fig:F2}(b) we plot the filter functions for all WDD
sequences with $n<32$ and $r=4$.  We see that {\em all} have the same
low-frequency rolloff, validating the assertion that the order of
error suppression in a WDD sequence is determined by $r$.  
The actual coherence error, $1-W$, associated with a given DD sequence
can be calculated by using the relevant power spectrum
$S_\beta(\omega)$ for the noise in Eq. (\ref{Eq:chi}). The presence of
spikes in the filter function at $\omega=O(\tau_{\text{min}}^{-1})$ (not shown)
implies that we can only expect error suppression as long as
$\omega_{c}\tau_{\text{min}}<1$.  We can guarantee this by shrinking
$\tau_{\text{min}}$ through use of more frequent decoupling pulses, as
far as the technological limitations on pulse rates permit this (see
Ref. \cite{Viola:BADD} for a discussion of the lower bounds on error
suppression in DD at constrained minimum switching time).

Fig.~\ref{Fig:F2}(c) depicts the performance of $\text{WDD}_{2^{r}-1}$
sequences for a specific illustrative case where a $1/\omega^{2}$
noise power spectral density $S(\omega)$ persists up to a Gaussian
high-frequency cutoff $\omega_{c}$ (given in units of inverse time
interval $1/\tau$). We have scaled the noise to have a value
comparable to that derived from the low-frequency behavior of nuclear
spin diffusion in singlet-triplet qubits~\cite{BiercukPRB2011}.  We
observe, as expected, that not only does the calculated $1/e$
coherence time increase with $n$, but the slope of the error
accumulation at short times also increases with $n$.  This is a
manifestation of the increasing order of error suppression with $r$
described above, and shows the utility of high-order WDD sequences for
quantum computing applications where minimizing the error probability
is of utmost importance~\cite{NC2000}.

\subsection{The Walsh sequence suite}
\label{subsec:classification}

It is clear from the previous discussion that a number of familiar
integer-based DD sequences can be identified as special instances of
WDD sequences. These sequences are described next for reference and
are summarized in Table~\ref{tab:Table1}.  For example, periodic DD
involves repetitive application of uniformly spaced $\pi$
pulses~\cite{Viola1998}. A Walsh function composed of a single
Rademacher function $R_{k}(x)$ ($r=1$) corresponds to a PDD with
$2^{r+1}-1$ pulses.  CPMG sequences are modifications of PDD in which
the first and last free-evolution periods are half the duration of the
interpulse period. We refer to a CPMG sequence with $n$ pulses as
$\text{CPMG}_{n}$. A Walsh function of the form $R_{k}(x)R_{k-1}(x)$,
with $k \geq 2$, corresponds to $r=2$ and $\text{CPMG}_{2^{k}}$.

We also observe that the sequences defined by $\text{WDD}_{2^{r}-1}$,
identified in the previous section, are in fact concatenated DD (CDD)
sequences. The latter are known to allow arbitrary orders of error suppression in
DD for arbitrary noise models by recursive embedding of a sequence in
a ``larger'' one \cite{Khodjasteh2005,Khodjasteh2007}. As long as a
first-order DD sequence can be implemented in a system, concatenating
it with itself results in higher orders of cancellation, at the
expenses of increased sequence length.  For a purely dephasing
environment as examined here, the natural first-order DD sequence (the
``base'' for concatenation) is the well-known spin-echo sequence,
characterized by two equal intervals separated by a $\pi$ pulse; each
higher level of concatenation corresponds then to embedding the spin
echo in a larger spin echo \cite{Cywinski2008,Hodgson2010}. Using
induction, we can show that each concatenation level corresponds to
multiplying the sequence propagator by a Rademacher function.  This
results in a product of all Rademacher functions up to order $r$, thus
$b_1=\ldots = b_r=1$, which corresponds to $\text{WDD}_{2^{r}-1}$. In
summary,
\begin{equation}
\text{CDD}_{r}\leftrightarrow\textsf{R}_{1}(x)\cdots\textsf{R}_{r}(x)
\leftrightarrow\text{WDD}_{2^{r}-1}.
\label{wcdd}
\end{equation}
\noindent 
Again, the WDD sequences that correspond to CDD have a propagator
corresponding to a product of Rademachers of all orders from $1$ to
$r$, giving $r$-order error suppression with the lowest value of
sequency, $s$.

\begin{table}[t]
\centering
\begin{tabular}{|c|c|c|c|}
\hline $n$ & $\text{WDD}_{n}$ & \parbox[t]{1in}{Number of Pulses\\
(Sequency)} & \parbox[t]{1in}{Rolloff\\ (dB/octave)}\tabularnewline
\hline \hline $2^{r}$ & PDD & $2^{r+1}-1$ &
$6\times(1+1)$\tabularnewline \hline $2^{r-1}+2^{r}$ & CPMG & $2^{r}$
& $6\times(2+1)$\tabularnewline \hline $2^{r}-1$ & CDD &
$\lceil\frac{2^{r+1}-2}{3}\rceil$ & $6\times(r+1)$\tabularnewline
\hline
\end{tabular}
\caption{Familiar sequences among WDD$_n$. $\lceil x\rceil$ denotes
the ceiling function, that is, the smallest integer not less than $x$.
The exact expression for the number of pulses in CDD was obtained with
the aid of \cite{oeis}.}
\label{tab:Table1}\end{table}

The WDD family significantly enlarges the sampling set of DD sequences
relative to the much more constrained CDD. If the minimum switching
time is constrained to $\tau_{\text{min}}=\tau/2^{m}$, then all the
$2^{m}$ $\text{WDD}_{n}$ sequences with $n=0,\cdots,2^{m}-1$ are
viable. In contrast, the subset of viable CDD sequences for the same
fixed total time contains only $m$ sequences, an exponentially smaller
number.  (However, it remains an even smaller subset of {\em all}
possible digital DD sequences, which contains $2^{2^{m}}$ sequences. This potential advantage will be
discussed in Sec. \ref{Sec:Benefits}.)  

WDD thus forms a unifying mathematical framework for the generation of digital DD sequences, including many familiar sequences and a large variety of novel sequences.  Using insights derived above, we may fully characterize the structure of arbitrary Walsh functions,
and therefore arbitrary WDD modulations.  By the structure of the
Walsh functions themselves, {\em all} WDD sequences can be produced
recursively from free evolution by combining the following two
intuitive operations:

$\bullet$ {\em Repetition}, where the sequence propagator $y(t)$ is
repeated identically to produce a longer sequence:
\begin{eqnarray*}
y(t) & \mapsto & 
\begin{cases}
y(t/2), & t<\tau/2 ,\\
y((t-\tau)/2), & \tau/2<t<\tau.
\end{cases}\\
\end{eqnarray*}

$\bullet$ {\em Concatenation}, where the sequence propagator $y(t)$ is
repeated with the sign reversed, still yielding a longer
sequence:
\begin{eqnarray*}
y(t) & \mapsto & 
\begin{cases}
y(t/2), & t<\tau/2,\\
-y((t-\tau)/2), & \tau/2<t<\tau .
\end{cases}
\end{eqnarray*}

The actual implementation of repetition and concatenation
may involve inserting a $\pi$ pulse in the middle of the formed
sequence to account for the required sign change at $t=\tau/2$.  In
general, $\text{WDD}_{2n}$ is constructed by repeating
$\text{WDD}_{n}$, whereas $\text{WDD}_{2n+1}$ is constructed by
concatenating $\text{WDD}_{n}$.  For instance,
$\text{WDD}_{30}=(\text{WDD}_{15})(\text{WDD}_{15})$ [repetition]
and
$\text{WDD}_{31}=(\text{WDD}_{15})\pi(\text{WDD}_{15})$ [concatenation],
but note the reversal of the role of the middle pulse in
$\text{WDD}_{14}=(\text{WDD}_{7})\pi(\text{WDD}_{7})$ [repetition]
or $\text{WDD}_{15}=(\text{WDD}_{7})(\text{WDD}_{7})$ [concatenation].

Each concatenation increases the rolloff slope by one order (by
contributing an additional sine factor in $F_{\text{WDD}_n}$), whereas
each repetition does not improve the rolloff (a cosine factor) and may
instead increase the bandwidth of the filter, $\Omega_{p}$.
This produces a diversity of design features that can be used to
improve coherence \emph{times} (e.g. the $1/e$ decay time denoted
$T_{2}$) in addition to coherence \emph{values} (or, equivalently,
error rates) by choosing the appropriate Walsh basis function.


\section{Beyond Classical Phase Noise and Quantum Memory}
\label{Sec:Beyond}

In this section, we consider different extensions of the idea of Walsh
modulation to control settings more general than examined thus far.

\subsection{Quantum phase noise}

Decoherence in a quantum system is most generically described by the
interactions with a \emph{quantum} environment.  We focus first, as
before, on a single qubit (see below for multi-qubit extensions). In
the absence of control and an internal system Hamiltonian, evolution
under the noise Hamiltonian of Eq. (\ref{eq:Hn}) is then replaced by
evolution under an open-system dephasing Hamiltonian of the form
$$H=H_{SB}+H_{B},\;\;\;H_{SB}=\sigma_{z}\otimes B_{z},$$ 
\noindent 
where physically $H_{SB}$ represents the interaction term, and $B_z$
and $H_{B}$ are generic operators acting on the environment,
respectively.  Typically, $H_{SB}$ causes entanglement between the
system and the environment, which eventually results in loss of phase
coherence and mixed qubit states.

The WDD sequences can be applied to the above general quantum
dephasing scenario.  Under a sequence of $\pi$ pulses, the evolution
is identically generated by a piecewise-constant Hamiltonian
$\pm\sigma_{z}\otimes B_{z}+H_{B}$.  Here, the filter function formalism is 
not exactly applicable in general.   An
important exception, as noted, is provided by the case of a bosonic
bath \cite{Viola1998,Uhrig2007,Hodgson2010}, and results accurate up
to the second order in $H_{SB}$ have been established in
\cite{uchiyama02} for an arbitrary quantum dephasing environment under
periodic DD (see also \cite{kitajima2010} for exact results on more
general classical noise models). Nonetheless, a description based on
an effective Hamiltonian and the Magnus expansion can be used to
approximate the evolution of the system provided that $\Vert
B_{z}\Vert$ (operator norm of $B_{z}$) is sufficiently small.  Making
note of the concatenated and repetitive structure of WDD and following
Ref.~\cite{Khodjasteh2007}, we can show that the norm of the effective
Hamiltonian for the evolution under $\text{WDD}_{n}$ is given by:
\begin{equation}
\Vert H_{\text{WDD}_{n}}\Vert=O[\Vert B_{z}\Vert \tau^{r}\max(\Vert
B_{z}\Vert,\Vert H_{B}\Vert)^{r}],
\label{eq:hwdd}
\end{equation}
where the strength of the bare interaction $\Vert B_{z}\Vert$ is
scaled by a factor proportional to $\tau^{r}$, with $r$ denoting, as
before, the Hamming weight of $n$. This implies that the expansion of
the propagator for the evolution in $\tau$ starts with the power
$\tau^{r+1}$. The fidelity loss can be defined as a distance between
the ideal state (original state in DD) and the actual propagated state
of the system at the end of the evolution (after tracing out the
environmental degrees of freedom)
\cite{lidarzanardikhodjasteh2008}. In the limit of small actions $\tau
H_{\text{WDD}_{n}}$, the fidelity loss scales with $\Vert \tau
H_{\text{WDD}_{n}}\Vert^2$ or, equivalently, {\em the qubit fidelity
loss scales at worst with $\tau^{2(r+1)}$}. While this result mirrors
the error cancellation properties of WDD in the filter function
formalism [Eq. (\ref{fwdd})], we emphasize that asymptotic
relationships such as Eq. (\ref{eq:hwdd}) must be understood as
scaling laws that may hide (possibly large) $r$-dependent
prefactors. This fact highlights the importance of Eq. (\ref{Eq:chi})
as an exact expression for Gaussian phase noise for which a
counterpart does not exist in a general quantum-dephasing setting.

\subsection{Generic decoherence}
\label{Generic}

As already mentioned, DD can be applied to a large class of open
quantum systems: a control propagator that regularly traverses the
elements of the so-called DD group in principle allows suppression of
arbitrary non-Markovian decoherence, including multi-qubit noise in
many-qubit systems \cite{Viola1999}.  When the noise operators are
decomposed into algebraically independent components, one may
reconstruct the generic decoupling procedure through concatenation as
well \cite{Khodjasteh2007}.  While the group-theoretic DD design is
applicable only to first-order error suppression, concatenation allows
us to combine DD sequences for different noise axes to produce a high
order \emph{and} generic decoupling procedure. This idea is the basis
of a variety of recent DD schemes, such as quadratic DD (QDD)
\cite{WestQDD_2010} and its
multi-qubit variants
\cite{Liu2011,Jiang2011}.

The WDD sequences can be similarly concatenated along different axes
to allow suppression of general noise. For example, consider a generalized
classical noise Hamiltonian acting on a qubit:
\[
H_{\text{noise}}=\beta_{x}(t)\sigma_{x}+\beta_{y}(t)\sigma_{y}+
\beta_{z}(t)\sigma_{z} ,\]
\noindent 
where $\beta_{i}(t)$ are stochastic processes. Applying a WDD sequence
with $\sigma_{x}$ pulses effectively removes the
$\beta_{y}(t)\sigma_{y}+\beta_{z}(t)\sigma_{z}$ noise component while
it leaves $\beta_{x}(t)\sigma_{x}$ intact.  Embedding this sequence
itself in a WDD sequence with $\sigma_{y}$ (or $\sigma_{z}$) pulses
removes the remaining $\beta_{x}(t)\sigma_{x}$ component as well. We
can generalize this idea to define {\em generic Walsh dynamical DD}
(GWDD) for a qubit. The control propagator $U_{c}(t)$ for the sequence
$\text{GWDD}_{n}$ is given by
\[
U_{c}(t)=\sigma_{x}^{[x(t)+1]/2}\sigma_{y}^{[y(t)+1]/2},
\]
where 
\begin{eqnarray*}
x(t) & = & \textsf{R}_{j_{1}}(t)\textsf{R}_{j_{3}}(t)\cdots,
\textsf{R}_{j_{2r-1}}(t),\\ y(t) & = &
\textsf{R}_{j_{2}}(t)\textsf{R}_{j_{4}}(t)\cdots\textsf{R}_{j_{2r}}(t),
\end{eqnarray*}
and, as in the single-axis case, the $\{j_{k}\}$ indices label the
locations of non-zero digits in the binary representation of $n$. In
practice, the final sequence corresponds to applying $\sigma_{x}$ or
$\sigma_{y}$ pulses at rates given by the the corresponding Rademacher
functions, taking into account the algebraic simplifications such as
$\sigma_{x}\sigma_{x}=I$ or $\sigma_{x}\sigma_{y}=\sigma_{z}$ and
ignoring all the resulting $\pm$ signs. We note that for $n=2^{2r}-1$,
$\text{GWDD}_{n}$ reproduces the generic CDD sequence of level $r$
described in Ref.  \cite{Khodjasteh2005}, while repetitions of
GWDD$_n$ include truncated periodic CDD protocols (so-called PCDD)
such as investigated in \cite{Wen2007,Wen2008}.  Again the Walsh functions form a unifying mathematical basis for sequence construction, reproducing familiar decoupling protocols.

For a multi-qubit system where qubits interact with the environment
{\em linearly} \cite{Viola1999,khodjasteh2009dcg}, GWDD is also
applicable, except that each unitary pulse operation has to be
replaced with a collective version that affects all qubits
simultaneously and equally. We remark in passing that the idea of
Rademacher products (concatenation in particular) can in principle be
extended to generic finite-dimensional control systems provided that
the role of $\sigma_{x}$ and $\sigma_{y}$ is replaced with the
generators of the corresponding DD group \cite{wocjan2001universal}.

\subsection{Non-identity operations}
\label{gate}

Our focus so far has been on error suppression while preserving
arbitrary quantum states, hence effectively implementing the identity
operator with higher fidelity than
free evolution for a given duration.  The periodic properties of the filter functions
associated with WDD sequences also allow us to achieve suppression of
frequency noise while effecting {\em non-trivial} (non-identity)
quantum gates, close in spirit to {\em dynamically corrected gates}
(DCGs)
\cite{khodjasteh2009dcg,khodjasteh2009pradcg,khodjasteh2010cdcg}.

The starting point in DCG constructions is the separation of the
action of a gate on a noisy system into ideal and error parts, both
represented as unitary operators.  Concretely, let $Q$ and $U_{Q}$
denote the ideal unitary gate that is to be implemented and the actual
unitary propagator corresponding to the evolution during the control
that aims to implement $Q$, respectively. We can write
\[U_{Q}=Q\exp(-iE_{Q}), \] 
\noindent 
where $E_{Q}$ is the error per gate (or {\em error action}) associated
with $Q$ and the goal is to minimize $E_{Q}$ for any desired $Q$. The
basic intuition is to use the separation of the ideal gate action and
the error to mix and match error parts so that altogether the errors
cancel out in a perturbative manner.

DCGs have algebraic connections to DD sequences but they remove the
need for instantaneous ideal pulses and are used to implement
non-identity unitary actions on the system.  The general theory for
constructing DCGs to higher orders for arbitrary systems appears in
\cite{khodjasteh2010cdcg}, but for our purpose we focus on an example
that applies to a recent experiment on correcting errors due to laser
frequency jitter in a multi-qubit entangling gate mediated by laser
light~\cite{Hayes2011}.  

Consider first the spin-echo sequence $XfXf$,
where $X$ denotes an ideal $\sigma_x$ gate and $f$ a free evolution
interval. The latter can be interpreted as a \emph{primitive}
implementation of the identity gate $I$. Thus, we may write the spin
echo as
\begin{equation}
I^{(1)}=XI^{(0)}XI^{(0)}, 
\label{Eq:DCGI}
\end{equation}
where $I^{(0)}$ refers to a zeroth-order approximation for the
identity action through free evolution, and $I^{(1)}$ refers instead
to an improved approximation of the identity action with a higher
order of error suppression. By recursively applying
Eq.~(\ref{Eq:DCGI}), or by repeating an improved identity gate, we
obtain the WDD sequences corresponding to progressively better and
longer approximations of the identity gate.  Consider next a {\em
specialized} scenario in which the $X$ gates are still ideal but,
instead of free evolution, we apply a target gate $Q$ with an
associated error $E_Q$. We also assume that the intended gate action
$Q$ and the $X$ gates commute. We begin with the sequence $XQXQ$ whose
actual propagator is given by
\begin{align}
& & XQ\exp(-iE_Q)XQ\exp(-iE_Q)= \nonumber \\
& & Q^2X\exp(-iE_Q)X\exp(-iE_Q),
\label{Eq:Q2}
\end{align}
where $X\exp(-iE_Q)X\exp(-iE_Q)$ can be interpreted as a spin-echo
sequence for free evolution with an effective Hamiltonian $E_Q$. Thus,
if $E_Q$ consists of terms that {\em anticommute} with $X$ (i.e. $X
E_Q X=-E_Q$), we expect that the sequence $XQXQ$ suppresses error in a
manner similar to $XIXI$, except that the resulting action will now be
the non-trivial gate $Q^2$. Furthermore, we may concatenate or repeat
this sequence to obtain higher-order suppression of errors at the
expense of a longer gate sequence.

The model which we just described for a simple high-order DCG applies
to the experimental setup in Ref. \cite{Hayes2011}, in which a
spin-motional entangling gate is applied to trapped ion qubits.  The
fundamental concept of this gate is that by state-selectively exciting
the harmonic-oscillator motional modes of ions in a shared trapping
potential, it is possible to entangle the internal spins of the
qubits. The actual unitary propagator associated with a single gate is
given by
\begin{align}
U_Q(t)=e^{S_N(\alpha(t)a^\dagger-\alpha^*(t)a)}Q,
\label{SMgate}
\end{align}
where $Q$ is a geometric phase gate, described above, that entangles
the qubits.  This gate relies on disentangling the spin and motion at
the end of the gate by detuning the driving force from a motional
resonance by $\delta$ and setting the gate time $t=j 2\pi/\delta$,
where $j$ is a positive integer.  In this case the drive and motion
desynchronize at $t$, and one precisely implements $Q$. Residual
spin-motional entanglement corresponds to an error and we thus define
the error per gate as
\begin{align*}
-i E_Q=S_N \,[\alpha(t)a^\dagger-\alpha^*(t)a].
\end{align*}
The coefficient $\alpha(t)$ is the time dependent displacement due to
the laser detuning $\delta$, that is, $\alpha(t)=(\Omega/2)\int_0^t ds
\exp(-i\delta s)$.  Such error may be due, in practice, to the fact
that $\delta$ typically carries an additive frequency error $\Delta$
for which we have $$ \alpha(t)=\frac{\Omega}{2}\int_0^t ds
\exp[-i(\delta+\Delta)s].$$ In this case we find that the harmonic
oscillator does not produce a ``closed loop'' in phase space, thus
yielding an error due to residual spin-motion entanglement.

Our goal is to achieve an improved approximation of the entangling
gate in Eq. (\ref{SMgate}) by cancellation of $\alpha(t)$ to
a high order at the end of the evolution. Besides varying the detuning
$\delta$, the evolution of $\alpha(t)$ can also be controlled by
flipping the phase of the laser-mediated optical dipole force used to
drive the ion motion (a nearly instantaneous action), which
effectively corresponds to switching the sign of the
interaction. Consider the evolution of the system punctuated by such
phase flips occuring at the times $p=\{t_j\}_{j=1}^n$, 
with $t_0=0$ and $t_{n+1}=\tau$ to denote the total gate/evolution
duration. The expression for $\alpha(\tau)$ is then given by
$$\alpha(\tau)=\frac{\Omega}{2}\sum_{j=0}^n (-1)^j
\int_{t_j}^{t_{j+1}} ds \exp[-i(\delta+\Delta)s], $$
and we note that
\begin{align*}
\int_{t_j}^{t_{j+1}}\! ds \exp[-i(\delta+\Delta)s]=
-i\frac{e^{-i(\delta+\Delta)t_j}-e^{-i(\delta+\Delta)t_{j+1}}}{\delta+\Delta}.
\end{align*}
In reality, the mismatch $\Delta$ may follow a probability
distribution $P(\Delta)$ and we can write the expected value of $\vert
\alpha(\tau)\vert^2$ as
\begin{align}
\overline{\vert\alpha(\tau)\vert^2
}=\frac{\Omega^2}{4}\int_{-\infty}^\infty
\frac{F_p\left((\delta+\Delta)\tau\right)}{(\delta+\Delta)^2} P(\Delta) d\Delta ,
\end{align}
where we have used the expression for the filter function in
Eqs. (\ref{eq:fff}) and (\ref{eq:ff}). This expression resembles the
equation for the decoherence error associated with a DD sequence
applied at the same times, except that the {\em argument of the filter
function in the integrand is shifted by a value $\delta$}, and instead
of the power spectrum $S_\beta(\omega)$ the error probability density
$P(\Delta)$ is used. To cancel the effects of $\Delta$ we require the
filter function to have a zero of high multiplicity at
$\omega=\delta$.

\begin{figure}[tb]
\begin{centering}
\includegraphics[width=\columnwidth]{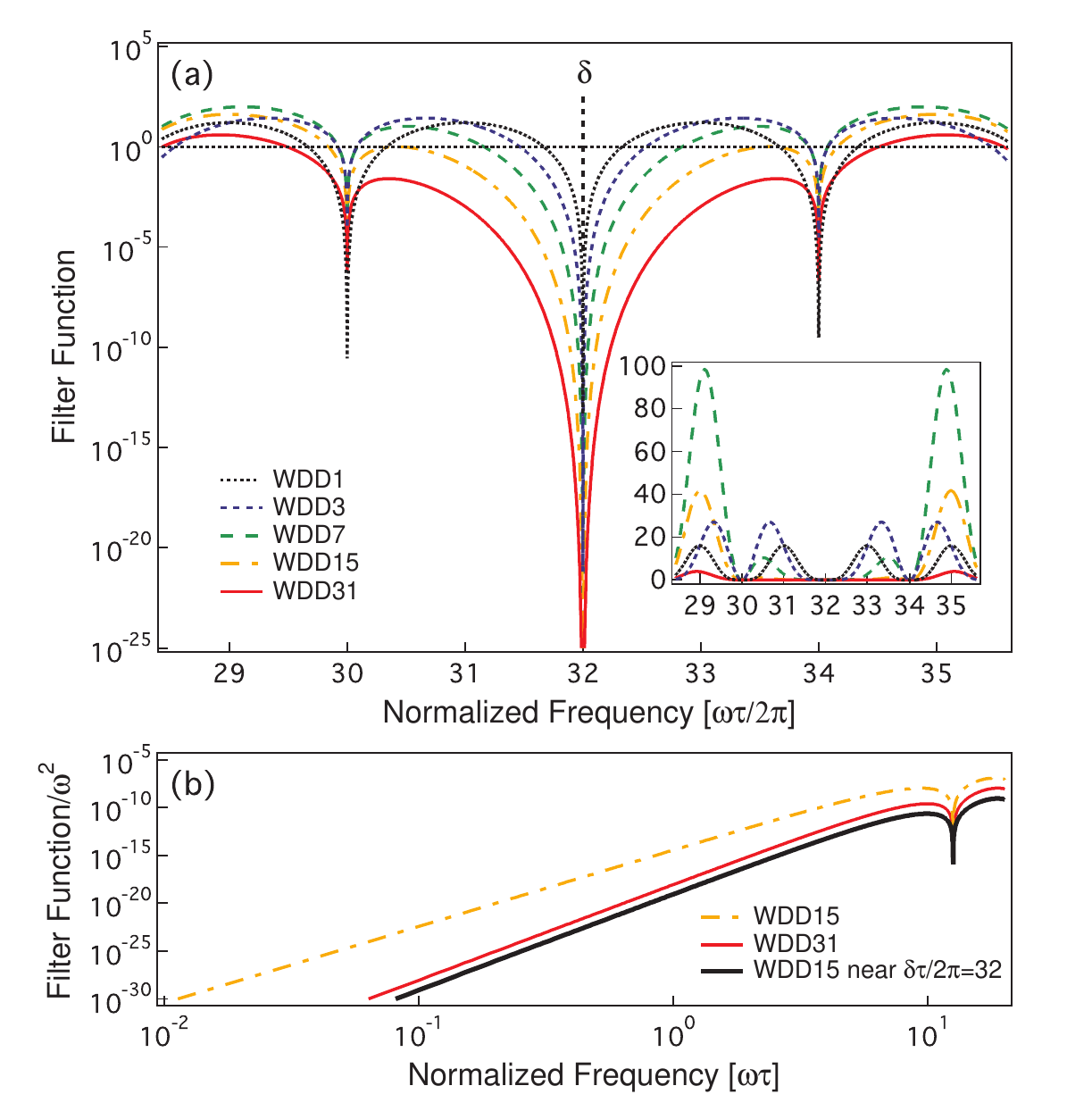}\\
\caption{(color online) Noise suppression at fixed frequency. (a) Filter function
about normalized frequency $\delta$ on a semilog scale for various WDD
sequences, resulting in a notch where noise is suppressed. The notch
bandwidth increases with sequency and the order of error suppression
about $\delta$ increases with $r$. Inset: Same filter functions
plotted on a linear scale. (b) Filter function normalized by
$\omega^{2}$ for WDD15 and WDD31. The solid thick line shows WDD15 at
$\delta/2\pi=2^{5}$, where the horizontal axis represents detuning
from $\delta$. Here the order of error suppression as a function of
detuning has increased by one order and is comparable to that for
WDD31 near zero frequency. See text for details.  \label{Fig:DCG} }
\par\end{centering} \centering
\end{figure}

Such a zero may be realized in the filter function of the WDD
sequence; from Sec. \ref{subsec:Rademacher}, we recall that the filter
function $F(\omega\tau)$ has a natural period (translational
invariance, see Fig. \ref{Fig:DCG}) at $\tau_{\text{min}}/2\pi$,
corresponding to $\tau/(2^{m+1}\pi)$. This produces the requisite
zeroes in the filter function at fixed values of $\delta$, a feature intrinsic to the formulation of WDD.  

In order to gain a quantitative understanding, let us calculate the Fourier transform of the sequence propagator, 
\begin{equation}
\tilde{y}(\omega\tau)=\int_{0}^{\tau}
dt\textsf{W}_{n}(t/\tau)e^{i(\delta+\Delta)t/\tau}.
\label{eq:filt:func:Delta}
\end{equation}
We extract the filter function's dependence on $\Delta$ near $\delta$ by use of an equality that is similar to
Eq. (\ref{eq:Rad:int}),
\begin{equation}
\int_{0}^{1}dx\prod_{\left\{ k\right\}
}\textsf{R}_{k}\left(x\right)e^{i2^{r+1}\pi
x}\sum_{i=0}^{r}a_{i}x^{i}=0,
\label{eq:second:Rad:int}
\end{equation}
where the set $\left\{ k\right\} $ has $r$ elements in it, $k\geq1$
and $\textrm{max}\left(k\right)\leq r$.  A sketch of the proof of
Eq. (\ref{eq:second:Rad:int}) appears in the Appendix.  If the
frequency $\delta/2\pi=2^{k}/\tau$, then
Eq. (\ref{eq:filt:func:Delta}) can be written as
\begin{equation}
\tilde{y}(\omega\tau)=\tau\int_{0}^{1}dt'\prod_{\left\{ k\right\}
}\textsf{R}\left(k,t'\right)e^{i2^{r+1}\pi
t'}\sum_{m=r+1}^{\infty}\frac{(i\Delta't')^{m}}{m!},
\end{equation}
where we have defined $t'=t/\tau$ and $\Delta'=\Delta\tau$. This
expression indicates that $\tilde{y} (\omega\tau)$ scales like
$\Delta^{r+1}$ in the vicinity of $\delta$. Thus, the slope of the
rolloff as a function of detuning is set by $r$, giving increasing
robustness to fluctuations in detuning from $\delta$.  For fixed
$r$, the bandwidth of the ``notch'' increases with sequency, as does
the equivalent $\Omega_{p}$. These properties are
expressed in Fig~\ref{Fig:DCG}.  

While the filter function is
translationally symmetric, we see that the form of filter performance
around $\delta$ is not the same as it is around $\omega=0$, due to the
factor of $\omega^{-2}$ appearing in Eq.~\ref{eq:fff}.
An expansion as a function of $\Delta$ therefore yields improved
performance relative to the zero-frequency rolloff when accounting for
this factor.

It thus follows that for the specific case treated here by flipping the laser phase at times
corresponding to the $\text{WDD}_n$, the error in the detuning
$\delta$ can be suppressed up to order $r+1$, where as before $r$ is the
Hamming weight of $n$.  This general approach was studied experimentally in
Ref.~\cite{Hayes2011}.  Interestingly, this sequence can be interpreted as a high-order DCG
where the primitive gates $Q$ are defined as $\tau_\text{min}$-long
periods of evolution at detuning $\delta=2\pi/\tau_\text{min}$ and the
$X$ gates are implemented by phase flips.
We can interpret all the resulting DCGs generated by WDD sequences as
repetitions and concatenations of the basic sequence $XQXQ$, resulting
in a corrected gate $Q^{2^m}$ which is the desired entangling
operation but has an error that scales with $\Delta^{2(r+1)}$.


\section{Benefits of WDD over other optimized approaches}
\label{Sec:Benefits}

Interest in the DD community has recently focused on optimized pulse
sequences producing high-order suppression of noise either through
analytical approaches (e.g. UDD~\cite{Uhrig2007}), numerical
optimization (e.g. Locally Optimized ODD (LODD)~\cite{Biercuk2009},
Optimized noise Filtration DD (OFDD)~\cite{Uys2009}, Bandwidth Adapted
DD (BADD)~\cite{Viola:BADD}), or combinations of optimization and
concatenation strategies (e.g. concatenated UDD
(CUDD)~\cite{Uhrig2008_2}, QDD~\cite{WestQDD_2010}, Nested UDD
(NUDD)~\cite{Liu2011}).  These sequences bring many benefits in terms
of resource-efficient DD, as they generally optimize error-suppression
against pulse number.

Studies have shown, however, that the extraordinary 
benefits provided by optimized sequences are largely suppressed in the
presence of realistic constraints such as imperfect control
pulses~\cite{BiercukPRA2009,Xiao2011,Wang2010}, digital
clocking~\cite{Biercuk_Filter}, and timing limitations
\cite{Hodgson2010}. Moreover, noise suppression benefits have been
shown to be minimal for $S_{\beta}(\omega)$ dominated by low-frequency
noise and exhibiting slowly decaying high-frequency
tails~\cite{Uhrig2008,Uys2009,Uhrig2010}.

The WDD sequences possess benefits over existing approaches along
two primary metrics: 

$\bullet$ Efficient hardware sequencing;

$\bullet$ Restricted search space for sequencing.

\noindent We will separately explore each of these benefits next.

\subsection{Efficient hardware sequencing}

Current experiments in DD employ a user-programmed microprocessor and
a complex hardware chain ({\em e.g.}, a signal generator controlled by
a programmable logic device responsible for pulse timing and under PC
control). This is appropriate for demonstration experiments, but fails
to provide a scalable solution due to both sequencing challenges and
the difficulty of input-output (I/O) in complex or quantum systems. We
are therefore interested in finding solutions permitting all sequence generation to be performed at the
\emph{local} level.  

Once we accept this consideration, our metrics of
efficiency change relative to the majority of published literature.  When considering optimized DD, sequencing is complex and will likely require either a local microprocessor to
decode instructions and apply a DD sequence, or multiple high-bandwidth communication pathways
to external controllers. Both situations pose challenges in terms of
significant local power dissipation in control hardware and energy
inflows associated with I/O pathways. The energy expended in the
number of pulses applied may contribute only a small amount of the
total local power dissipation given these considerations. Given a presumed need for local DD sequence generation we thus arrive at digital sequencing complexity as the relevant metric for efficiency.

WDD meets this challenge, its primary benefit being that the Walsh
functions are easily produced using simple digital
circuits~\cite{Walsh_Beauchamp}. We identify Rademacher function
generators (square-wave generators) as basic hardware resources for
the physical-layer implementation of dynamical error suppression,
realizing the control propagator in real-time. Rademacher functions
may be generated in hardware with relative ease from a distributed
clock signal, and may be added/multiplied together via hardware logic
to generate Walsh toggling frames (for instance Harmuth's array
generator~\cite{Walsh_Beauchamp}).  The number of Rademacher functions
required to achieve a given error suppression may therefore be deemed
a relevant quantitative means of establishing the sequencing
complexity. 

As we have shown, a WDD sequence derived from $r$ Rademacher functions
cancels the first $r-1$ orders of dephasing noise, and the sequences
derived from $\textsf{W}_{2^{r}-1}(x)\equiv$ CDD$_{r}$ do so for the
smallest value of sequency (for fixed $r$). WDD therefore provides the
highest-possible order of error suppression with respect to the number
of Rademacher functions required to implement a given DD protocol,
providing efficiency in the sequencing complexity. For the same
sequencing complexity $r$, using sequences with different $n$ permits
modification of $\Omega_{p}$, giving flexible control over noise
filtering capabilities.  By contrast, other optimized sequences discussed above require substantially greater sequencing resources, as the Walsh-transformation of a sequence such as UDD indicates the need for a large $r$ in order to generate the requisite $\{\delta_{j}\}$.

Furthermore, the structure of WDD and Walsh
function generation is compatible with a nested or concatenated
structure of decoupling about multiple axes, which enables the
generation of GWDD sequences for generic decoherence.  In the special
case of a large separation of timescales for different decoherence
processes, GWDD also allows for efficient bit-stacking of Walsh
functions derived from similar hardware, but executed with different
time-bases.  Finally, we note that certain sequences constructed by
recursively repeating the same base sequence many times may find value
in circumstances where long-time storage and low-latency memory
interrupts are desired in addition to improved low-frequency error
suppression.  This topic will be addressed in a separate manuscript.

Once a given Walsh function (or set of Walsh functions) is generated,
conversion to WDD in real time may be achieved via hardware
differentiation or edge-triggering of separate circuitry producing
pre-programmed control pulses. Overall complexity is reduced by separating
\emph{control sequencing} from \emph{pulse generation}.

Using Walsh functions for generation of the control propagator and
hardware techniques for the triggering of applied pulses therefore
provides a means to efficiently realize dynamical error suppression at
the local level. Through this approach the need for access to a
user-controlled microprocessor in order to send pulse-sequence
commands to hardware, or for a local processor to interpret externally
generated commands may be obviated: all control and timing may be
performed using relatively simple digital logic realized in an
Application Specific Integrated Circuit.  A system protected by WDD
sequences may run semi-autonomously, implementing pre-selected WDD
sequences, or it may be programmed externally, where the only
communicated information required could be the value of $n$ in binary
(or as Gray code), a repetition number, and a trigger signal starting
the sequence.

A significant challenge in such schemes is suppression of spurious rising/falling edge signals
arising from, say, propagation delays in digital circuitry. However,
latching WDD sequences to the clock signal should aid suppression of
these signals. Accounting for nonzero-duration control pulses with $\tau_{\pi}$ an integer multiple of the clock period, may also be achieved through use of latching logic to introduce hardware delays while the control operations are applied.  However, the effect of such delays on WDD performance would need to be characterized in detail.  The creation of an optimized \emph{hardware-based} generator of WDD sequences remains a problem for future study.

\subsection{Restricted search space}

While there already exists a plethora of DD procedures within the
bang-bang limit, in practice not all these procedures will be suitable
to the operating range of physical systems or the demands of quantum
information processing protocols.  This is true despite the fact that
the objective in all such procedures is identical: maximize the
fidelity of preserving an arbitrary quantum state for a desired
time. For sufficiently simple experimental systems, an empirical
optimization procedure (such as LODD) can sample over the space of all
permissible DD pulse sequences and search for an optimal one.  This
idea has also been explored numerically in the above-mentioned OFDD
and BADD protocols, as well in optimized DD for power-law spectra
\cite{Uhrig2010}.  These approaches grow significantly in complexity
as the pulse number increases, for instance in a large scale system
requiring long-term storage.  The central idea in WDD -- that
sequences using digital timing are made by attaching or repeating
smaller sequences recursively -- can be instrumental in significantly
reducing the search space for finding optimal DD sequences empirically
or numerically.  We briefly outline potential implications of this
idea here.

The canonical model of single-qubit dephasing with a given noise power
spectral density used through most of this paper retains considerable
structure in choosing when to apply the $X$ pulses.  This persists
even when enforcing digital timing conditions as below.  Given
$N=2^{m}$ time-bins, each of duration $\tau_{\text{min}}$,
one may elect to either apply or not apply a pulse, thus generating a
space of $2^{N}=2^{2^{m}}=2^{2^{\tau/\tau_{\text{min}}}}$ possible
sequences. Performing a complete search over this space becomes
extremely challenging as the total duration $\tau$ of the DD procedure
grows, as observed in numerical approaches to generating randomized DD
protocols \cite{Viola2006,Santos2008} and to optimized sequences
\cite{Biercuk2009,Uys2009}. 

WDD provides a natural means to reduce the search space for the
generation of dynamical error suppression sequences as there are only
$N=2^{m}=2^{\tau/\tau_{\text{min}}}$ WDD sequences within the
operational constraints of the problem.  It also provides an intuitive
analytical framework allowing pre-selection of certain sequences
within WDD, further reducing the search space.  For instance, a search
might exclude all WDD sequences with $r<r_{\text{min}}$, as they will
be known to provide insufficient low-frequency noise suppression.
With such constraints, and the relative simplicity of implementing WDD
in hardware, these sequences become especially attractive for an
empirical search in which the performance of each sequence is tested
in an actual experimental setting.

Perhaps even more interesting are the options offered when Walsh DD is
extended to generic decoherence of a larger set of qubits in a quantum
memory. Recently, an efficient perturbative DD procedure for this task
has been explored (NUDD) that utilizes $(D_{S}^{2})^{r}$ pulses for
$r$th order decoupling of a generic $D_{S}=2^{n_{q}}$-dimensional
system of $n_{q}$ qubits \cite{Liu2011}. This procedure assumes the
most general form of errors, including many-body errors.  However, in
reality error models are generally sparse. A fundamental problem of DD
of many qubit systems (closely related to quantum simulation
algorithms
\cite{wocjan2001universal,leung2002simulation,rotteler2008dynamical})
is to optimize the decoupling when the error model for the interaction
of the qubits is locally sparse. Using the procedure described in
Sec. \ref{Generic}, we can progressively produce a search space for
such an optimization procedure starting from elements of a basic DD
sequence and recursively building longer sequences by the WDD
construction procedure (repetition/concatenation).


\section{Conclusion}

In this manuscript we have studied the problem of reducing sequencing complexity in a dynamical error suppression framework, through consideration of \emph{digital} modulation schemes.  Towards this end we have introduced the Walsh functions as a
mathematical basis for the generation of dynamical error suppression
sequences.  We have revealed how Walsh dynamical decoupling naturally incorporates familiar sequences
(e.g. PDD, CPMG, CDD), and examined the properties of
all the possible recursively structured sequences corresponding to
Walsh functions.

Our analysis has demonstrated that the order of error suppression
achieved by a given WDD sequence is set by the number of elementary
Rademacher functions required to generate that sequence.  This is
manifested as a scaling of the low-frequency rolloff in the filter
function as $6(r+1)$ dB/octave, with $r$ the number of Rademacher
functions appearing in the sequence. Meanwhile, the high-frequency
performance of a WDD sequence, and hence its noise-suppression bandwidth, is tunable via selection of $n$ for a given $r$, providing significant flexibility in sequence construction.
Further we have introduced a simple means to construct WDD sequences
using concatenation and repetition, and a technique to suppress
general errors via GWDD.

We have shown that the Walsh functions provide efficient performance
against control complexity, quantified by the value of $r$, and the
supporting sequencing hardware. We believe that sequencing complexity
will serve as a useful new metric with significant weight in
system-level analyses, going far beyond standard optimization over
pulse number in a DD sequence.  

These considerations are likely to make the Walsh functions an
attractive framework for the development of a quantum memory
incorporating hardware-efficient physical-layer error-suppression
strategies.  Interestingly, the problem of searching for (digital) DD
sequences with high order error suppression properties is also related
to finding Littewood complex polynomials with high order zeroes
\cite{Newman1988,Golan2006}, which may point to further connections
with both signal processing theory and polynomial analysis and
approximation. From a practical standpoint, the variety of WDD
sequences realizable through simple hardware sequencing provides a
flexible solution for many future experimental systems.

\appendix

\section{Order of error suppression}

In this section we prove that for a given Paley ordering $r$, as long
as $i\le r$:
\begin{equation}
\int_{0}^{1}\textsf{R}_{1}(x)\cdots \textsf{R}_{j_{r+1}}(x)x^i dx=0, 
\label{eq:prove}
\end{equation}
where $j_k$ are non-zero positive integers in increasing order. This
identity can be used to show the error suppression of the WDD
sequences {\em directly in the time domain} and can be mathematically
interpreted as the vanishing of the \emph{moments} of the Walsh
functions on the unit interval.

We proceed by induction on $r$ starting with the base case, $r=0$:
\begin{equation}
\int_{0}^{1}\textsf{R}_{j_{1}}(x)dx=0 . 
\label{eq:base}
\end{equation}
Since $j_{1}>0$, the corresponding Rademacher function
$\textsf{R}_{j_{1}}$ will be periodic and balanced on $[0,1]$, which
validates the base case in Eq. (\ref{eq:base}).

For the inductive step, let us assume that 
\begin{equation}
\int_{0}^{1}\textsf{R}_{j_{1}}(x)...
\textsf{R}_{j_{r+1}}(x)x^{i}dx=0,
\label{eq:indas}
\end{equation}
where $i\le r$. We need to prove that
\begin{equation}
\int_{0}^{1}\textsf{R}_{j_{1}}\left(x\right)...
\textsf{R}_{j_{r+1}}(x)\textsf{R}_{j_{r+2}}(x)
x^{i}dx=0,\label{eq:indres}
\end{equation}
where $i\le r+1$. 

The sign of Rademacher function $\textsf{R}_{j}(x)$ is determined by
the $j$-th binary digit of $x$ which we refer to as $b_{j}(x)$. When
$i=0$, Eq.~(\ref{eq:indres}) reduces to the average (zeroth moment) of
the product of Rademacher functions over the interval $[0,1]$. We give
a probabilistic argument for the cancellation of this average which
can also be proven using induction. The average of the product of the
Rademacher functions can be written as
\begin{equation}
\int_{0}^{1}(-1)^{b_{j_1}(x)}\cdots(-1)^{b_{j_{r+2}}(x)}
dx .
\label{eq:average}
\end{equation}
We can interpret the above integral as an expectation value (denoted
below by $E$) of a function of a real random variable $x$ distributed
uniformly over the interval $[0,1]$. The corresponding (Lebesgue)
probability measure $dx$ is equivalent to products of {\em
independent} and uniform (discrete) measures for each binary digit of
$x$ \cite{Adams}.  We can thus rewrite Eq. (\ref{eq:average}) in the
form
\begin{equation*}
E[(-1)^{b_{j_1}}\cdots (-1)^{b_{j_{r+2}}}]=E[(-1)^{b_{j_1}}]\cdots
E[(-1)^{b_{j_{r+2}}}]=0.
\end{equation*}
  
We can then use the $i=0$ case as the base for another induction proof
over $i$, where we assume that Eq. (\ref{eq:indres}) holds for all
$i'\le r$. We refer to this as the ``inner induction'' assumption and
proceed to prove Eq. (\ref{eq:indres}) for $i=r+1$, which we rewrite
as
\begin{equation}
I=\int_{0}^{1}\textsf{W}_{n}(x)\textsf{R}_{j_1}
\left(x\right)x^{r+1}dx. 
\label{eq:return}
\end{equation}
Note that we have singled out the lowest order Rademacher function in
the product and written the product of the rest as a Walsh function
with Paley ordering $n$. We can rewrite the integral as
\begin{equation*}
I=\Big(\int_{0}^{2^{-j_1}}\!-\int_{2^{-j_1}}^{2\times 2^{-j_1}}\!+
\int_{2\times 2^{-j_1}}^{3\times 2^{-j_1}}\!-\cdots \Big)
\textsf{W}_{n}(x)x^{r+1}dx,
\end{equation*}
where the integrals match the positive and negative values of the
Rademacher function $\textsf{R}_{j_1}$.
\begin{widetext} 
Making the substitution $x'=x-2^{-j_1}$ for the negative-signed terms,
we can write
\begin{equation}
I=\Big(\int_{0}^{2^{-j_1}}\!+ \int_{2\times 2^{-j_1}}^{3\times
2^{-j_1}}\!+\cdots
\Big)[x^{r+1}\textsf{W}_{n}(x)-(x+2^{-j_1})^{r+1}\textsf{W}_{n}
(x+2^{-j_1})]dx.
\end{equation}
The Walsh function $\textsf{W}_n$ is constructed entirely of
Rademacher functions with indices larger than $j_1$ having periods
that commensurate with $2^{-j_1}$, which results in
$\textsf{W}_{n}(x+2^{-j_1})=\textsf{W}_{n}(x)$. This allows us to
write
\begin{equation}
I=\Big(\int_{0}^{2^{-j_1}}\!+ \int_{2\times 2^{-j_1}}^{3\times
2^{-j_1}}\!+\cdots \Big)\textsf{W}_{n}(x)[x^{r+1}-
(x+2^{-j_1})^{r+1}]dx. 
\end{equation}
\end{widetext}
We can rewrite the above as an integral over $[0,1]$ using the kernel
$(\textsf{R}_{j_1}(x)+1)/2$:
\begin{equation*}
I=\int_{0}^{1}\textsf{W}_{n}(x)\frac{\textsf{R}_{j_1}
(x)+1}{2}[x^{r+1}-(x+2^{-j_1+1})^{r+1}]dx ,
\end{equation*}
where the leading $x^{r+1}$ powers cancel when we expand
$(x+2^{-j_1+1})^{r+1}$, leaving us with an integrand in which every
power of $x$ appears with an exponent $i'$ less than or equal to $r$.
This matches the assumptions of our inner induction, leading to the
cancellation of the integral for $i={r+1}$ and thus establishing the
main result in Eq. (\ref{eq:prove}).

We also note the following result, related to the WDD filter function
at higher frequencies and used in Sec. \ref{gate} of the main text:
\begin{equation}
\int_{0}^{1}\textsf{R}_{1}(x)\cdots \textsf{R}_{j_{r}}(x)e^{i 2^{r+1}
\pi x} x^i dx=0,
\end{equation}
where $i\le r$ and we have used the highest Rademacher function in the
product with a trigonometric function of the same period. This result
can also be proven by induction in a similar manner. 

It is interesting to note that the only properties of the Rademacher
functions that are used in our proof are their ``frequencies'' and
symmetry.  Thus, any family of functions that mimics the periods and
the sign changes 
of the Rademacher function family could in principle be used in our
proof.

\begin{acknowledgments}
This work was partially supported by the US Army Research Office under
Contract Number W911NF-11-1-0068, the Australian Research Council
Centre of Excellence for Engineered Quantum Systems CE110001013, and
the US National Science Foundation under Award PHY-0903727 (to LV).
This research was partially funded by the Office of the Director of
National Intelligence (ODNI), Intelligence Advanced Research Projects
Activity (IARPA), through the Army Research Office. All statements of
fact, opinion or conclusions contained herein are those of the authors
and should not be construed as representing the official views or
policies of IARPA, the ODNI, or the U.S. Government.
\end{acknowledgments}

\bibliography{Walshbib}

\begin{thebibliography}{80}
\bibitem[{not({\natexlab{a}})}]{note0}
\bibinfo{note}{Note that when $s$ is odd, a final pulse must be added just
  before the readout to reset the qubit state.}
\bibitem[{not({\natexlab{b}})}]{note2}
\bibinfo{note}{If $s$ and $n$ are stored in digital registers, we have
  \texttt{s=n\textasciicircum{}(n\textgreater{}\textgreater{}1)} where
  \texttt{\textasciicircum{}} denotes bitwise addition mod 2 (XOR) and
  \texttt{\textgreater{}\textgreater{}} denotes a shift right operation.}
\end{thebibliography}
 
\end{document}